\documentclass[10pt,twocolumn,showkeys,showpacs,preprintnumbers,prd,superscriptaddress,nofootinbib]{revtex4-1}
\usepackage{verbatim}
\usepackage[T1]{fontenc}
\usepackage[utf8]{inputenc}
\usepackage[american]{babel}
\usepackage{epsfig}
\usepackage{booktabs}
\usepackage{multirow}
\usepackage{dcolumn}
\usepackage{amsmath}
\usepackage{mathtools}
\usepackage{ragged2e}
\usepackage{amsfonts}
\usepackage{amssymb}
\usepackage{float}
\usepackage{ulem}
\usepackage{epstopdf}
\usepackage{bm}
\usepackage{siunitx}
\usepackage{braket}
\usepackage{enumitem}
\usepackage{soul}
\usepackage[table]{xcolor}
\usepackage{color}
\usepackage{transparent}
\usepackage{pifont}
\usepackage{enumitem}
\usepackage{orcidlink}
\newcommand{\software}[1]{%
  \vspace{0.5em}
  \noindent\textit{Software:} #1
}

\definecolor{deeppink}{rgb}{1.0, 0.08, 0.58}
\definecolor{purple}{rgb}{128, 0, 128}
\definecolor{blueblue}{RGB}{0,112,255}

\hypersetup{
    colorlinks=true,
    linkcolor=deeppink,
    citecolor=deeppink,
    urlcolor=blueblue}

\usepackage{hyperref}

\begin{document}

\title{
First tomographic measurements of the angular clustering and bias of photometric quasars from S-PLUS}

\author{Maria Lopes\orcidlink{0000-0001-9181-5675}}
\email{marialopes@on.br}
\affiliation{Observatório Nacional, Rua General José Cristino 77, 
São Cristóvão, 20921-400 Rio de Janeiro, RJ, Brazil}

\author{Felipe Avila\orcidlink{0000-0002-0562-2541}}
\email{felipeavila@on.br}
\affiliation{Observatório Nacional, Rua General José Cristino 77, 
São Cristóvão, 20921-400 Rio de Janeiro, RJ, Brazil}

\author{Armando Bernui\orcidlink{0000-0003-3034-0762}}
\email{bernui@on.br}
\affiliation{Observatório Nacional, Rua General José Cristino 77, 
São Cristóvão, 20921-400 Rio de Janeiro, RJ, Brazil}

\author{Lilianne Nakazono\orcidlink{0000-0001-6480-1155}}
\email{liliannenakazono@on.br}

\affiliation{Observatório Nacional, Rua General José Cristino 77,
São Cristóvão, 20921-400 Rio de Janeiro, RJ, Brazil}

\affiliation{Departamento de Física Matemática, Instituto de Física,
Universidade de São Paulo, R. do Matão 1371,
05508-090, São Paulo, SP, Brazil}

\begin{abstract}

We investigate the quasar clustering in the QuCatS photometric catalog, constructed from the fourth data release of the Southern Photometric Local Universe Survey (S-PLUS). 
The quasar sample for this study consists of $23,402$ quasars, spanning the redshift interval $1.1 \leq z_{\rm phot} \leq 2.6$, which we divided into four bins for our tomographic analyses. 
In each bin, we measured the two-point angular correlation function and compared it with theoretical predictions for the clustering of dark matter in the flat-$\Lambda$CDM cosmological model. 
We then estimate the effective linear bias in each bin that best fits the data, obtaining $b_{Q}(z_{\rm eff}=1.26)=2.17^{+0.30}_{-0.35}$, $b_{Q}(z_{\rm eff}=1.63)=2.67^{+0.36}_{-0.42}$, $b_{Q}(z_{\rm eff}=2.04)=3.29^{+0.52}_{-0.62}$, and $b_{Q}(z_{\rm eff}=2.38)=4.05^{+0.71}_{-0.86}$. 
These measurements are consistent with previous analyses in the literature. 
Moreover, the evolution of the bias agrees well with the $b(z)$ parameterization proposed by \cite{laurent2017}. 
Our measurements constitute the first cosmological study of the tomographic clustering of photometric quasars provided by the S-PLUS dataset, contributing to the mapping of the evolution of linear bias and paving the way for future analyses of data releases with larger datasets. 
In addition, our study also serves as a consistency test for the calibration procedure of the photometric redshift probability distributions of the quasars in the QuCatS catalog.

\end{abstract}



\maketitle


\section{Introduction}
\label{sec:introduction}

The clustering analysis of cosmological tracers is one of the main tools to study the large-scale structure (LSS) of the universe, which relates the observed spatial distribution of these objects to the theoretical predictions of structure formation and evolution \cite{peebles1980, padmanabhan1993, coles1995}. 
Mapping the clustering evolution of these tracers across cosmic epochs allows us to understand when and how the oldest structures in the Local Universe formed from the gravitational growth of small density fluctuations present in the early universe \cite{Ross2009}, as well as their relationship to the underlying dark matter. 
In addition to these studies, various statistical analyses of cosmological observables have been used to investigate properties of the LSS, including studies of statistical isotropy and homogeneity~\citep{Lopes2024, Lopes2025, mokeddem2025, Aluri2023, McConville2023}.

In this context, quasars, correctly classified as an extragalactic object 
by~\cite{Schmidt1963}, 
are active galactic nuclei whose emission is produced by the accretion of matter onto a supermassive black hole (SMBH) at the center of the host galaxy \citep{Salpeter1964, Lynden-Bell1969, Krolik1999}. 
They are so luminous that they can be detected at high redshifts, with recent measurements exceeding $z > 7$ \cite{Yang2026, Banados2018, Wang2021}, and are considered the most energetic sources in the universe. 
Furthermore, the high amplitude observed in the quasar two-point correlation function indicates that these objects reside preferentially in massive dark matter halos (e.g., \cite{Croom2001}) and, therefore, act as biased tracers of the underlying dark matter field, as predicted for objects inhabiting high-density regions \citep{Kaiser1984, Shen2009}. This characteristic makes quasars valuable tracers of the LSS, as they provide a map of the dark matter field at cosmic epochs.

In the last two decades, several studies have investigated the clustering features of quasars, showing that the bias increases with redshift, whether through direct measurements of the bias parameter $b$ by fitting the clustering amplitude to angular \citep{Myers2006, Myers2007b, DiPompeo2014, Donoso2014, DiPompeo2015, DiPompeo2017, Myers2007, Ross2009, Timlin2018}, or in the redshift space 
two-point correlation functions~\citep{Croom2001, arita2023, Ivashchenko2010, Shen2009, laurent2017, Laurent2016, Eftekharzadeh2015, Croom2005}. 
%
In addition, some analyses consider cross-correlation measurements with gravitational lensing of the Cosmic Microwave Background (CMB) \citep{Sherwin2012, Geach2013, DiPompeo2014b, DiPompeo2015, DiPompeo2017, Han2019, Geach2019, Piccirilli2024}, correlations with other tracers~\citep{Ikeda2015, He2017, Petter2022}, or even measurements of $b \sigma_8$~\citep{Porciani2006}. 
These studies encompass a wide variety of samples, including spectroscopic surveys such as the 2dF QSO Redshift Survey~\citep{Croom:2004eg} and SDSS/eBOSS~\citep{SDSS:2003rmd, York2000, Dawson2013}, as well as photometric quasar samples, yet they converge on a consistent scenario in which quasar bias increases monotonically with redshift. The agreement between clustering-based and CMB lensing-based bias estimates is especially relevant, as both methods are subject to independent systematic effects, providing strong support for the robustness of current quasar bias measurements~\citep{Belsunce2025}. 
Clustering measurements also provide constraints on cosmological parameters and primordial non-Gaussianity \citep{Alonso2023, Adame2025}, establishing a direct relation between the statistical properties of primordial fluctuations and the large-scale distribution of matter today.

In this work, we study the clustering of photometric quasars at intermediate-to-high $z$ from the QuCatS catalog, constructed from the fourth data release of the Southern Photometric
Local Universe Survey (S-PLUS)~\citep{nakazono2024}. We focus on a tomographic analysis, in which we measure the two-point angular correlation function in four tomographic bins of photometric redshift in the range $1.1 \leq z \leq 2.6$, and estimate the effective linear bias in each interval by comparing the observed angular correlation function with that of matter (with $b(z)=b_m=1$) predicted by 
the standard cosmological model, the flat-$\Lambda$CDM model. 
We then compare our results with spectroscopic and photometric measurements available in the literature over the past two decades.

The tomographic study in four disjoint redshift intervals is suitable for investigating the angular clustering of photometric redshift samples at different epochs of the universe evolution. 
For this task, we use the two-point angular correlation function, an approach that 
turns our analyses weakly dependent on a cosmological model~\citep{ulisses2026}. 

Thus, in this work we present the first cosmological study of quasar angular clustering from the photometric survey S-PLUS, measuring the evolution of photometric quasar bias and allowing a comparison with previous results. 

This work is organized as follows: In Section~\ref{sec:data}, we describe the observational, random, and simulated datasets used in our analysis. In Section~\ref{sec:methodology}, we present how we calculated the two-point angular and theoretical correlation functions, how we propagated the photometric redshift error, and how we adjusted the effective linear bias. In Section~\ref{sec:Results}, we present and discuss our results. Finally, we summarize our conclusions and discuss prospects in Section~\ref{sec:summary}.

\section{Datasets}\label{sec:data}

In this section, we detail the procedures used to select the quasar sample for analysis, and 
also the construction of the random catalog (designed to reproduce the observational properties 
and sky coverage of the dataset). 
Additionally, we describe the production of the mock catalogs used to calculate the covariance matrix. 

\subsection{QuCatS data}\label{sec:qucats}

In this study, we used the Quasar Catalog for S-PLUS (QuCatS)\footnote{The catalog is available for direct download at:
\url{https://splus.cloud/files/QuCatS_Nakazono_and_Valenca_2024.csv}}, compiled by \cite{nakazono2024}, and derived from the fourth data release (DR4) of the Southern Photometric Local Universe Survey (S-PLUS) \citep{Herpich2024}, 
as the survey’s first catalog of quasar candidates with reliable photometric redshifts. 
S-PLUS is a wide-field optical imaging survey that maps the southern hemisphere sky in 12 optical bands, using the Javalambre system \citep{Mendes2019} and the T80-South telescope, located in Chile at the Cerro Tololo Inter-American Observatory. 
S-PLUS provides catalogs of stars, galaxies, and quasars~\citep{Nakazono2021}, 
contributing to the study of astronomical objects at cosmic scales.

\begin{figure}[htbp]
    \centering
    \includegraphics[width=\columnwidth]{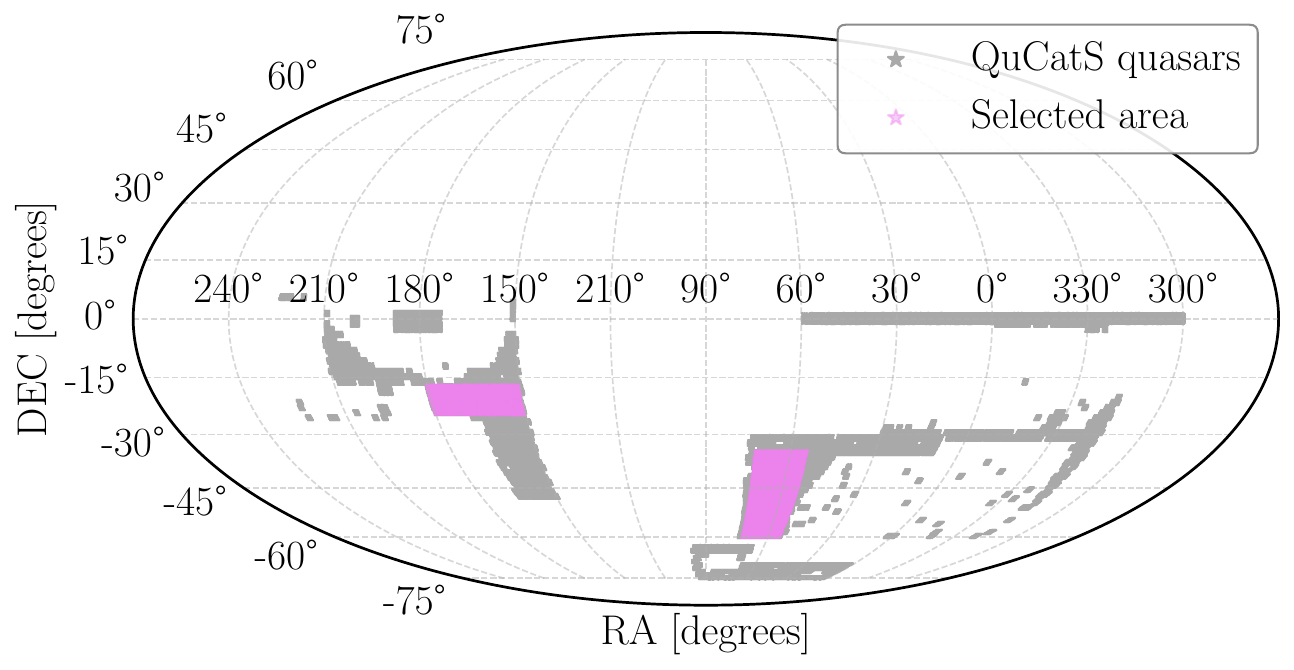}
    \caption{Distribution of the full QuCatS quasars from the S-PLUS survey (gray) and the selected analysis area of the quasar sample (violet) on the celestial sphere in Equatorial coordinates, shown in the Mollweide projection.}
    \label{fig:footprint}
\end{figure}

QuCatS is a value-added catalog that compiles $645,980$ quasar candidates with good photometric quality, and classification probabilities greater than $80\%$,  with magnitudes in $r$ band $16.95 < r <21.3$, comprehending most of the surveyed area, covering approximately 3,000 deg$^2$ of the southern sky, with the exception of the galactic disk region, which has high stellar density and high interstellar extinction, making quasar detection more challenging. The coverage area of these data is shown in gray color in Figure~\ref{fig:footprint}.

For each source, the catalog provides photometric redshift estimates obtained using different machine learning algorithms, such as Random Forest (RF)~\citep{Breiman:2001hzm}, Flexible Conditional Density Estimation (FlexCoDE)~\citep{10.1214/17-EJS1302}, and Bayesian Mixture Density Network (BMDN)~\citep{Bishop94,bishop_1997}, as well as the average of these estimates and their respective standard deviations. 
While Random Forest directly provides a single redshift estimate, the FlexCoDE and BMDN methods estimate the Probability Density Function (PDF) of the redshift. 
Thus, the catalog provides the complete PDFs for each quasar, represented by 200 discrete values in the case of FlexCoDE and by the parameters of a mixture of seven Gaussians in the case of BMDN. In addition, information on equatorial coordinates, classification probabilities, magnitudes, and other astrophysical properties is also available.

\subsection{Sample Selection}\label{sample_selection}

To obtain a suitable sample for analyzing the angular clustering of quasars, the sample used in this study was constructed from the QuCatS by adopting the following additional selection criteria:

(i) When selecting the final area for our study, we avoided regions with highly fragmented coverage, and also discarded regions near the Magellanic Clouds (or the Galactic plane) due to potential residual stellar contamination. 
After this search, we selected for analysis two rectangular regions that cover the main contiguous areas of the QuCatS surveyed area. 
This choice allows us to construct a random catalog that adequately reproduces the effective geometry of the regions in analysis, in addition to simplifying its computational generation, because S-PLUS did not provide sky masks to produce mocks. 
The final selected area, i.e. two rectangular regions, can be seen in Figure~\ref{fig:footprint} emphasized in deep pink color. 

(ii) We selected only high-confidence quasar candidates, that is, those with a probability of being classified as quasars greater than $95\%$ ($\texttt{PROB\_QSO} \geq 0.95$). Although the QuCatS consists of candidates with probabilities greater than $80\%$, adopting a more restrictive threshold reduces potential contamination from stars and galaxies, resulting in a accurate sample. 

\begin{figure}[t]
\centering
    
\includegraphics[width=\columnwidth]{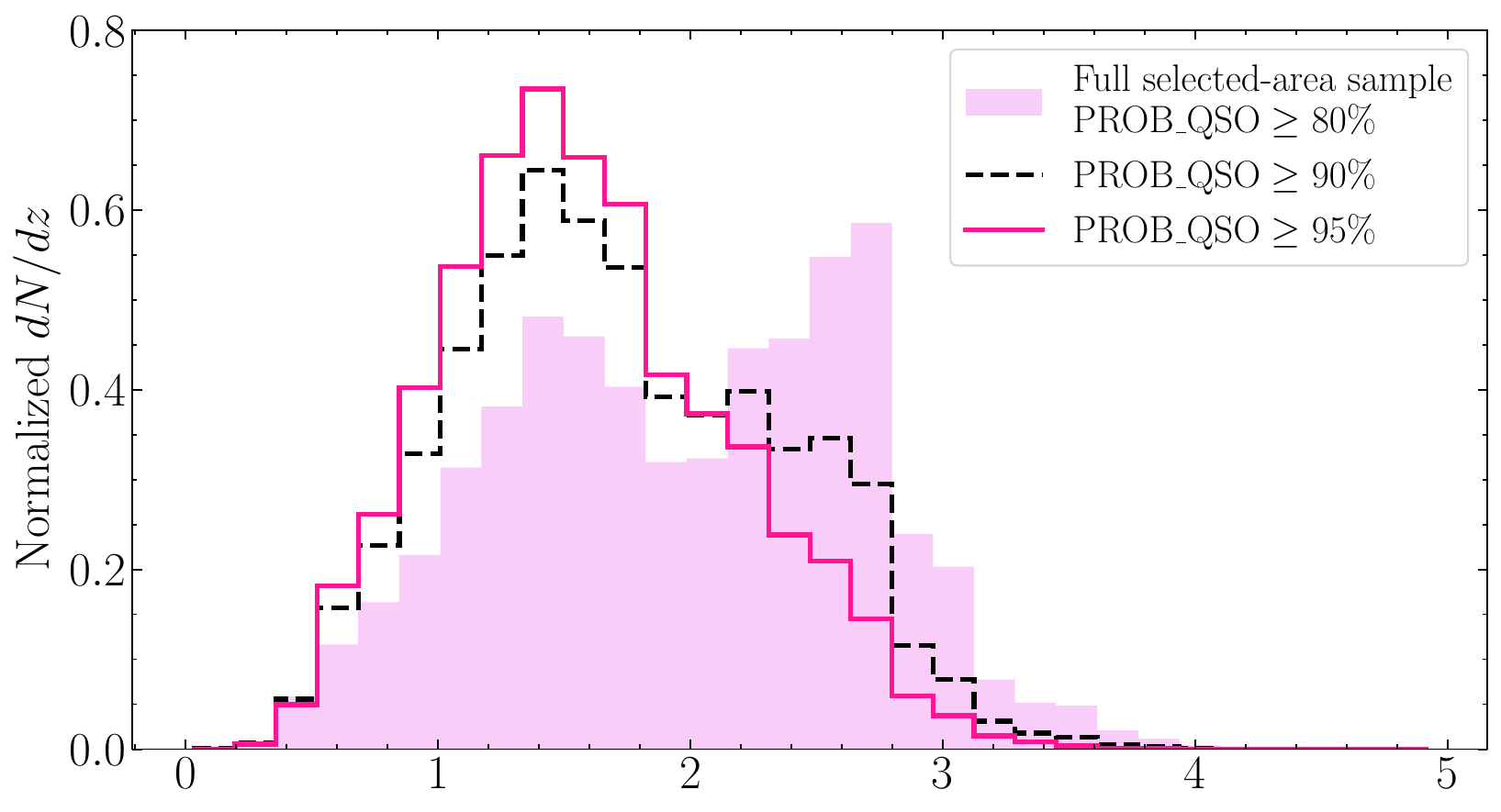}
\caption{Normalized redshift distribution, $dN/dz$, obtained by stacking the individual FlexCoDE photometric redshift probability density functions. The filled violet distribution corresponds to the full selected-area sample, while the black dashed and deep pink distributions correspond to the sub-samples with probability of being classified as quasar  greater than $90\%$ and $95\%$, respectively. All the stacked distributions are rebinned into 30 redshift bins for visualization purposes.} 
\label{fig:dndz}
\end{figure}

Regarding the second criterion, our choice was based on the behavior of the RF classification algorithm documented by \citep{Nakazono2021}, which was used to construct QuCatS. In that study, the authors used S-PLUS DR2 data and found that quasar classification exhibited the lowest accuracy (purity) compared to stars and galaxies in the sample, even when including data from the Wide-field Infrared Survey Explorer. 
The primary reason for this is the classifier's confusion between stars and quasars at the faint magnitude limit ($r \approx 22$); it can misclassify stars as quasar candidates, thereby contaminating the sample.
This contamination directly impacts the determination of quasar linear bias in clustering analyzes, 
since the inclusion of non-quasar objects systematically alters the contrast of density fluctuations, resulting in a dilution of the clustering signal.

In Figure~\ref{fig:dndz}, we show the photometric redshift distribution for sources within the area selected for analysis (shown in violet color in Figure~\ref{fig:footprint}) based on classification probability thresholds: the complete sample from the selected area (violet), $90\%$ (dashed black), and $95\%$ (deep pink). 
We observe that different classification probability cuts affect sample selection as a function of redshift, consequently, the object counts calculated across the redshift range do not reflect a uniform reduction of the original population. 

This happens because the resulting sample is governed by a selection function where the purity criterion is related to the observational properties of the objects, such as magnitude and color, 
primarily the detection of redshifted emission lines in the narrow S-PLUS bands, 
leading to the non-uniform completeness observed across the entire redshift interval. 

It is observed that the most restrictive cut adopted in our final analysis, i.e. ${\rm PROB}\_{\rm QSO} \geq 95\%$, yields a more well defined redshift distribution, concentrated around $z \simeq 1.55$, as previously indicated by \citep{nakazono2024}. 
Furthermore, the authors report that for high confidence quasars, the various photometric redshift estimation methods (RF, FlexCoDE, BMDN) tend to converge toward very similar profiles. Although this cut drastically reduces the sample's number density, it results in a selection with a lower contribution from potential stellar contaminants and a more well-defined radial window. 
Therefore, the probability threshold we adopt for sample selection, i.e. $95 \%$, provides a more reliable catalog of quasars for our analyzes.

In this work, we use the photometric redshift PDFs estimated by FlexCoDE to incorporate redshift uncertainties into the tomographic clustering analysis, which is discussed in Section \ref{sec:sampling}, because FlexCode with narrow-band presents the best Conditional Density Estimation (CDE) loss and with best general performance. 
Thus, after applying all the selected criteria, the sample consists of 32,636 objects, distributed across the entire redshift range covered by the catalog, 
i.e. $0.034 \leq z \leq 4.913$.

The redshift distribution is characterized by the dN/dz distribution, estimated by  stacking the normalized FlexCoDE photometric redshift PDFs of all sources. The stacked distribution is given by
\begin{equation}
\frac{dN}{dz} =
\sum_{i=1}^{N} p_i(z) \,,
\label{eq:dndz_stack}
\end{equation}
where $p_i(z)$ is the normalized photometric redshift PDF of the $i$-th quasar and $N$ is the number of objects in the sample in study. 
The resulting distribution was subsequently normalized to unit integral, 
\begin{equation}
\left(\frac{dN}{dz}\right)_{\rm norm} \equiv 
\frac{dN/dz}
{\displaystyle\int_{z_{\rm min}}^{z_{\rm max}} (dN/dz) \, dz} \,.
\label{eq:dndz_norm}
\end{equation}
The normalized resulting $dN/dz$ is shown in Figure~\ref{fig:dndz}, in deep pink color. 
This distribution was adopted as the input redshift distribution for the generation of the mock catalogs described in Section~\ref{sec:mocks}.

Our analyses consider four tomographic redshift intervals: $1.1 \leq z_{\rm phot} < 1.4$, $1.4 \leq z_{\rm phot} < 1.9$, $1.9 \leq z_{\rm phot} < 2.2$, and $2.2 \leq z_{\rm phot} \leq 2.6$, totalizing $23,402$ quasars in the whole redshift range 
$1.1 \le z_{\rm phot} \le 2.6$.

\subsection{Random catalog}\label{sec:random}

The random catalog covers the same observational region as the quasar sample and contains approximately 100 times as many objects, that is, $N_{\rm rand}=100$ $N_{\rm data}$, so that the contribution of Poisson noise associated with the random pairs is significantly smaller than that from the observational sample. 
For each rectangular region (see Figure~\ref{fig:footprint}) the right ascension, denoted by $\alpha$, was generated from a uniform distribution over the observed range, $\alpha_{\rm rand}\sim U(\alpha_{\min},\alpha_{\max})$, while the declination, denoted by $\delta$, was obtained from a uniform distribution in $u\sim U(\sin\delta_{\min},\sin\delta_{\max})$~\citep{Franco:2023rhd}. 
Finally, the random catalogs generated for each region were combined into a 
single catalog used to estimate the angular correlation function 
$\omega(\theta)$, which will be discussed in Section~\ref{sec:f-correlção}.


\subsection{Log-normal Mock simulations}\label{sec:mocks}

Simulated catalogs are important tools for estimating the statistical uncertainties of clustering measurements, as they reproduce the statistical properties of the large-scale distribution of matter tracers in astronomical surveys. In this work, we employ log-normal simulated catalogs~\citep{coles1991}, which are widely used in analyzes of large-scale structures because they provide an efficient approximation of the evolved matter density field while preserving its key statistical properties~\citep{xavier2016, avila2024a, Lippich2019}. Consequently, they have become a standard approach for constructing mock catalogs that describe well --on average-- the matter density field in the observed universe~\citep{agrawal2017,Favole:2020ywr,Ramirez-Perez:2021cpq}. 

To generate these realizations, we used the \texttt{GLASS} (Generator for Large Scale Structure) package\footnote{\url{https://glass.readthedocs.io/stable/index.html}}~\citep{tessore2023}, a publicly available open-source Python framework, that efficiently produces log-normal cosmological simulated catalogs, and integrates continuously into the analysis workflow developed in this work. 
The linear power spectrum of matter was calculated using \texttt{CAMB} (Code for Anisotropies in the Microwave Background)\footnote{\url{https://camb.readthedocs.io/en/latest/\#}}~\citep{camb2000, camb2002}, assuming a fiducial cosmology, used as input for GLASS to generate the log-normal realizations that reproduce the expected angular clustering signal.

Configuring the simulations for our analyses requires specifying the cosmological parameters of the reference cosmology and the parameters that describe the observational setup of the survey. 
The fiducial cosmology chosen to construct the simulated models is the flat-$\Lambda$CDM model, based on the results of Planck 2018~\citep{planck2020}, whose parameters are shown in Table~\ref{table:glass}. 
The bias function, $b(z)$, describes the relationship between quasar density fluctuations and the underlying matter density field. 
It is worth noting that the simulated catalogs in this work were generated assuming an evolving quasar bias, rather than a constant bias value. 
Since our tomographic slices span redshift shells of considerable thickness, 
i.e. $\delta z = 0.3 - 0.5$, the evolution of the quasar bias cannot be neglected, since the clustering scale of the tracer varies with redshift~\citep{Einasto:2022grd}. 
Incorporating this feature into the simulations therefore provides a more realistic description of the quasar density field. 
For our analysis in the redshift interval $1.1 \leq z \leq 2.6$, we adopt the redshift-dependent parameterization proposed by \cite{laurent2017}: $b(z)=0.278[(1+z)^2-6.565]+2.393$. Although their relation was originally calibrated over the range $0.9 \leq z \leq 2.2$, it covers the majority of the interval explored here, making it a reasonable approximation for the evolution of the quasar bias in our simulations.

The survey mask was generated using the publicly available
\texttt{Healpy/HEALPix}\footnote{\url{https://healpy.readthedocs.io/en/latest/}}
pixelization algorithm \citep{gorski2005, Zonca2019}, ensuring that the mock catalogs reproduce the angular survey footprint.
To reproduce the radial selection of the quasar sample, we used the normalized redshift distribution from the complete catalog (Equation~\ref{eq:dndz_norm}) to calculate the fractional contribution of objects within each radial slice adopted in the GLASS simulations. 
It is worth noting that the information on photometric dispersion is already encoded in the redshift distribution itself, which is used as input in the form of PDFs. 
Thus, it is not necessary to introduce an additional uncertainty 
$\sigma_0$ into the code GLASS, because if doing so results in a double-counting of the uncertainty associated with the photo-$z$, information already considered in the PDFs. 
The expected number density in each redshift shell was then obtained by multiplying these fractions by the average surface density of the observed sample, calculated from the total number of quasars and the effective area of the survey. 
The complete simulated catalogs were generated using this radial selection function. 
Finally, the same tomographic redshift intervals adopted for the observational sample were considered in the simulated catalogs, producing the corresponding simulated tomographic realizations used in the analysis.

In this approach, we generated $N_{\rm m} = 1,000$ mock catalogs, providing a statistically representative ensemble to estimate the uncertainties associated with the clustering measurements. 
The two-point angular correlation function was computed for each mock realization using the same methodology adopted for the observational sample. The resulting set of correlation functions was then used to estimate the covariance matrix of the measurements.

\begin{table}[H]
\caption{GLASS parameters used for mock realizations.}\label{table:glass}
\centering
\begin{tabular}{c}
\hline
\hline
\textbf{Survey configuration}\\
\hline
$z_{\rm grid}=[0.0,\,5.0],\ \Delta z=0.1$\\
$b(z)=0.278\left[(1+z)^2-6.565\right]+2.393$\\
$n_{\rm corr}=3$\\
$N_{\rm side}=1024$\\
$\ell_{\rm max}=1024$\\
\hline
\hline
\textbf{Cosmological parameters}\\
\hline
$\Omega_b h^2 = 0.02236$\\
$\Omega_c h^2 = 0.1202$\\
$h = 0.6727$\\
\hline
\end{tabular}

\begin{minipage}{\columnwidth}
\vspace{0.2cm}

\justify
\textbf{Note.} Survey configuration: $z_{grid}$ is the grid used to construct the redshift shells; $b(z)$ is the bias function, which we adopted the redshift-dependent parametrization proposed by \citep{laurent2017}; $n_{corr}$ is number of correlated radial fields retained during the field generation GLASS, $N_{side}$ is the HEALPix pixelization parameter; $\ell_{max}$ is the highest-order multipole used in angular power spectrum. 
Cosmological parameters were adopted from \cite{planck2020} (TT,TE,EE+lowE): $\Omega_b h^2$ and $\Omega_c h^2$ are the baryon and cold dark matter physical densities, and $h$ is the dimensionless Hubble parameter. 
These parameters were used to generate the set of $1,000$ log-normal mock catalogs.
\end{minipage}

\end{table}

\section{Methodology}\label{sec:methodology}

In this section, we describe our methodology for deriving the bias evolution of the selected quasar sample from QuCatS, calculating the two-point angular correlation function, propagating the photometric redshift uncertainties into the clustering measurements, and computing the covariance matrix.
 

\subsection{Two-point angular correlation function estimator}\label{sec:f-correlção}

The two-point angular correlation function (2PACF), $\omega(\theta)$, describes the angular clustering of tracers in the large-scale structure by quantifying the excess probability of finding pairs of objects separated by an angle $\theta$ in the sky relative to a random distribution \citep{peebles1980}.

In our analysis we employ the Landy–Szalay (LS) estimator~\citep{Landy1993}, 
widely used in the literature for clustering 
analysis~\citep{Avila2021a, Avila2021b, Franco2025} 
because it minimizes the variance of the estimate of the two-point correlation function, while reducing the bias associated with the survey geometry and finite-sample effects~\citep{Kerscher2000A}. 
The LS estimator is defined as
\begin{equation}
    \omega (\theta) \equiv
    \frac{DD(\theta) - 2\,DR(\theta) + RR(\theta)}{RR(\theta)}
    \label{eq:2pacf},
\end{equation}
where $DD(\theta)$, $RR(\theta)$, and $DR(\theta)$ denote the normalized quasar-quasar, random-random, and data-random pair counts, respectively, with the pair counts normalized by the total effective number of pairs in each catalog. 
The angular separation $\theta$ is given by
\begin{equation}
    \theta = \cos{}^{-1}\\\Bigl[
        \sin\delta_1\,\sin\delta_2
        + \cos\delta_1\,\cos\delta_2\,\cos(\alpha_1 - \alpha_2)
    \Bigr].
\end{equation}

In this work, the calculations above were performed using the publicly available \texttt{TreeCorr} package\footnote{\url{https://github.com/rmjarvis/TreeCorr}}~\citep{jarvis2004}, with $\text{bin}$\_$\text{slop} = \text{angle}\_\text{slop} = 0.01$ parameters.


\subsection{Photometric Redshift Inverse Sampling and Tomographic Binning}\label{sec:sampling}

In clustering analyzes based on photometric data, photometric redshifts are essential for defining the redshift intervals that determine the radial shells onto which the quasars are projected. The width of the redshift bins must be significantly larger than the mean photometric redshift uncertainty, i.e., $\langle \sigma_z \rangle / \Delta z \ll 1$, since this condition minimizes the migration of objects between adjacent bins due to redshift uncertainties, thereby reducing potential systematic effects in the analysis, such as incorrect pair counts in 2PACF \citep{ulisses2026}. 

However, each object in our sample is characterized only by the PDF generated with FlexCode; as previously mentioned, we do not have a point value with an associated uncertainty for the photo-z that would allow us to directly quantify the relationship between the bin and the photometric error. 
For this reason, to consistently incorporate the probabilistic information present in the PDFs, we adopted the inverse transform sampling method on the PDFs, generating Monte Carlo realizations of the redshifts and defining the tomographic bins based on these samples. 
This procedure was suitably adopted in the 2PACF analyses for the S-PLUS 
blue galaxies~\citep{ulisses2026, Franco2025b}. 
The methodology can be described as: 

i) For each quasar, FlexCoDE provides an estimate of the photometric redshift probability density function (PDF), $p(z)$, sampled at 200 discrete points, corresponding to the columns \texttt{z\_flex\_pdf\_[1--200]}. These values represent the probability densities evaluated at 200 uniformly spaced redshift values over the interval $0.034 \leq z \leq 4.913$. Thus, the PDF of each object is represented by the discrete set $p(z_1), p(z_2), \ldots, p(z_{200})$, which is subsequently normalized according to
\begin{equation}
p_{\rm norm}(z)=
\frac{p(z)}
{\displaystyle\int_{z_{\rm min}}^{z_{\rm max}} p(z')\,dz'} \,.
\label{eq:pdf_norm}
\end{equation}

ii) Each PDF is numerically integrated to obtain its cumulative distribution function (CDF), $F(z)$,
\begin{equation}
F(z)=\int_{z_{\rm min}}^{z} p(z')\,dz',
\end{equation}
which is then inverted. Next, a random number $u\in[0,1]$, drawn from a uniform distribution, is associated with a photometric redshift value obtained via the transformation
\begin{equation}
z = F^{-1}(u) \,.
\end{equation}
This procedure is repeated for each quasar in $N_{\text{MC}} = 1,000$ Monte Carlo simulations, resulting in the same number of sampled catalogs.

iii) For each sampled catalog, we divided the data into the four adopted tomographic bins and computed the corresponding 2PACF using Equation~\ref{eq:2pacf}. This procedure was repeated for all Monte Carlo realizations, and the final angular correlation function for each tomographic bin was obtained by averaging the 2PACF over all realizations at each angular separation.

The selection of tomographic bins was performed entirely via inverse transformation sampling of the complete FlexCoDE PDFs, avoiding the use of single-point estimators such as the peaks provided by the catalog, i.e z$\_$flex$\_$peak, which may not adequately capture multimodal PDFs~\citep{nakazono2024}. 
To validate the procedure, the expected number of objects in each bin was estimated by summing the individual probabilities of belonging to the bin, obtained by integrating the PDFs, and compared with the distribution of the number of objects resulting from the Monte Carlo simulations, as shown in Table \ref{table:nobjects}. 
A good agreement was observed between both estimates. 
To avoid regions in the redshift histogram where the photo-z model showed extrapolation, we first considered only objects in the interval $0.5 < z_{phot} < 4.2$. After a series of tests, we restricted our final analysis to the redshift range $1.1 \leq z_{phot} \leq 2.6$, with an estimated number of 23,402 objects in that range\footnote{Although the sample initially comprises $32,636$ quasars in the range $0.034 \leq z \leq 4.913$, as mentioned in Section~\ref{sample_selection}, sampling from the individual redshift PDFs results in an average of $23,402$ quasars within the selected redshift interval $1.1 \leq z \leq 2.6$, since objects may fall either inside or outside this redshift range across different realizations.}, 
which was subdivided into four tomographic bins with bin edges: $z_{\rm phot} = [1.1, 1.4, 1.9, 2.2, 2.6]$ (see Table~\ref{table:nobjects}). 
The definition of these intervals was based on a balance between the number of objects in each tomographic bin and the typical uncertainties in the phot-z estimates (i.e., FlexCode with narrow bands has $\sigma_{\text{NMAD}} = 0.039$). Among the configurations evaluated, this division gave the best performance in fitting the model to the data, resulting in the lowest values $\chi^2$ and was therefore adopted in the final analysis.

In addition, following previous photometric quasar clustering analyzes~\citep{Donoso2014,DiPompeo2014,Timlin2018,William2025}, our 2PACF measurements were taken at 11 logarithmically spaced angular separation bins in the range $0.05^{\circ} < \theta < 1^{\circ} $. This interval represents a compromise between preserving sufficient pair statistics at small angular separations, where statistical uncertainty increases due to the smaller number of pairs, and restricting the analysis to scales at which the clustering signal remains robust.

\begin{figure*}
\centering
\includegraphics[width=\linewidth]{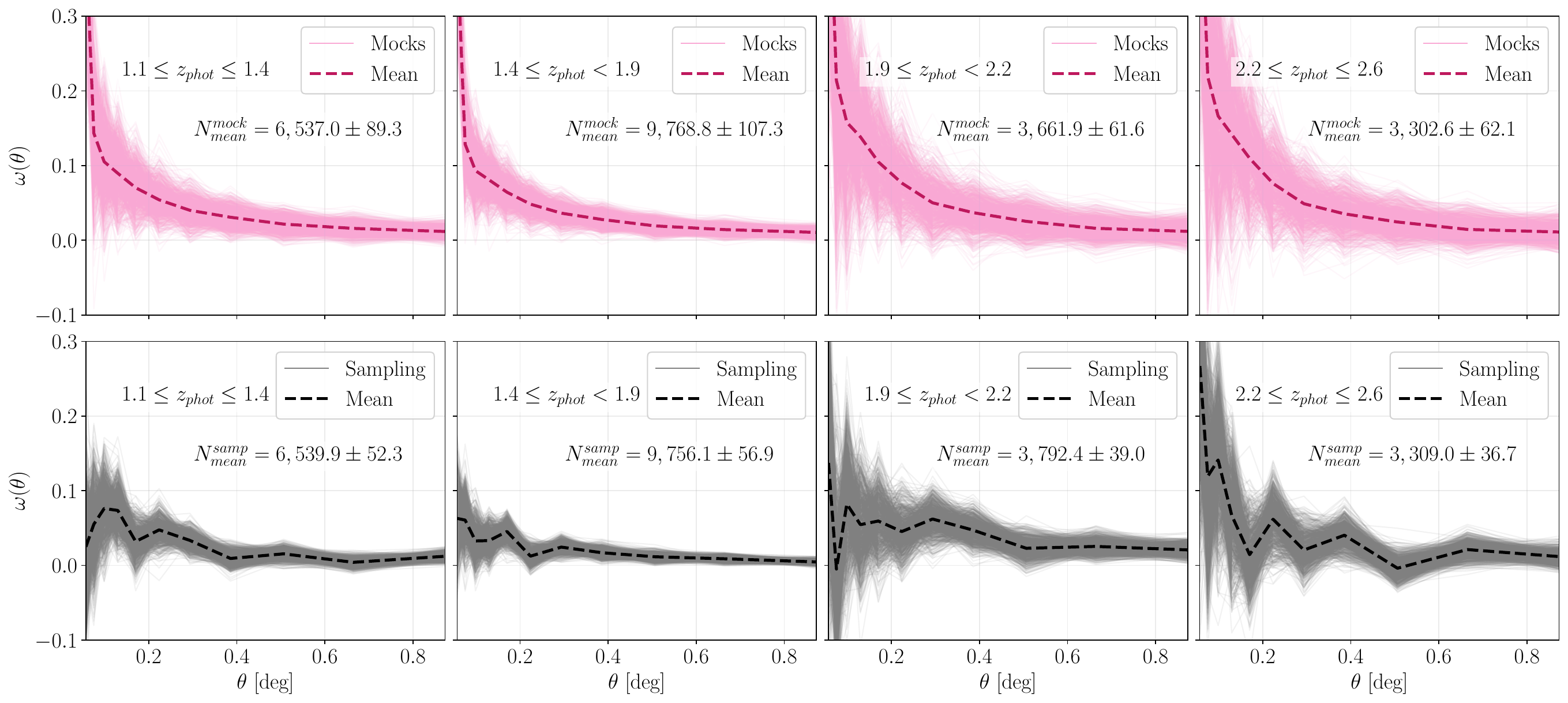}
\caption{Comparison of the angular clustering measured from the $N_{\rm m} = 1,000$ log-normal mock catalogs (upper panels) and from the $N_{\text{MC}} = 1,000$ photometric-redshift samplings (lower panels) for the four tomographic redshift bins. Individual realizations are shown as solid lines, while the dashed lines correspond to the mean 2PACF in each case. The panels, from left to right, represent the redshift intervals $1.1 \leq z_{\rm phot} < 1.4$, $1.4\leq z_{\rm phot}<1.9$, $1.9\leq z_{\rm phot}<2.2$, and $2.2\leq z_{\rm phot}\leq2.6$.
}
\label{fig:mocks_samplings}
\end{figure*}

The sampling results are shown in Figure~\ref{fig:mocks_samplings}. In the lower panel, the black dashed lines represent the average 2PACF in each tomographic bin, while the gray lines show the different samplings. The upper panel presents the equivalent result for the mocks (described in Section~\ref{sec:mocks}), where the deep pink dashed lines indicate the average 2PACF and the light deep pink lines correspond to the different realizations of the mocks.

\begin{table*}[htbp]
\caption{Expected number of objects per tomographic bin and comparison with Monte Carlo realizations.}
\centering

\begin{tabular*}{0.75\linewidth}{@{\extracolsep{\fill}}cccc}
\hline\hline
Interval & $N_{\mathrm{exp}}$ & $N^{samp}_{mean}$ & $N^{mock}_{mean}$\\
\hline
$1.1 \leq z < 1.4$ & $6,539.9 \pm 51.2$ & $6,539.3\pm{52.3}$ & $6,537.0 \pm{89.3}$\\
$1.4 \leq z < 1.9$ & $9,757.7\pm 56.6$ & $9,756.1\pm{56.9}$ & $9,768.8\pm{107.3}$\\
$1.9 \leq z < 2.2$ & $3,794.8\pm 39.5$ & $3,792.4\pm{39.0}$ & $3,661.9\pm{61.6}$\\
$2.2 \leq z \leq 2.6$ & $3,309.9\pm 35.5$ & $3,309.0\pm{36.7}$ & $3,302.6\pm{62.1}$\\
\hline
\end{tabular*}

\begin{minipage}{\linewidth}
\vspace{0.2cm}
\justify
\textbf{Note.} The expected number of objects $N_{\mathrm{exp}}$ in each tomographic bin is computed by integrating individual photometric redshift probability density functions using the cumulative distribution function evaluated at the bin boundaries, compared with inverse-sampling and mock realizations, confirming the consistency of the sampling method.

\end{minipage}

\label{table:nobjects}
\end{table*}

\subsection{Covariance estimation}\label{sec:cov_matrix}

Estimating the quasar bias requires the full covariance matrix of the measured 2PACF, which quantifies the correlations among the angular separation intervals.
The angular correlation function itself is measured from Monte Carlo realizations obtained via inverse-transform sampling of individual PDFs (as described in Section~\ref{sec:sampling}), thereby propagating photometric redshift uncertainties into the clustering measurement.

To estimate the covariance matrix in this study, we derive it from $N_{\rm m}=1,000$ log-normal mock realizations, generated as described in Section~\ref{sec:mocks}, which reproduce the survey geometry and incorporate the effective redshift distribution inferred from the same photometric redshift PDFs. These mock realizations therefore provide a self-consistent description of statistical uncertainties, including cosmic variance, shot noise, survey geometry, and mode coupling, such that no additional covariance term is included. 
Therefore, the covariance matrix is computed as
\begin{equation}
\mathrm{C}_{ij} =
\frac{1}{N_{\rm m}-1}
\sum_{k=1}^{N_{\rm m}}
[
\omega^{k}(\theta_i)-\bar{\omega}(\theta_i)
]
[
\omega^{k}(\theta_j)-\bar{\omega}(\theta_j)
],
\end{equation}
where $N_{\rm m}$ is the total number of mock realizations, $\omega_i^{(k)}$ is the value of the 2PACF measured in the $i$-th angular bin of the $k$-th mock; $\bar{\omega}(\theta_i)$ and $\bar{\omega}(\theta_j)$ are the mean value at the bin $i$ and $j$,
\begin{equation}
    \bar{\omega}(\theta_i)
    =
    \frac{1}{N_{\rm m}}
    \sum_{k=1}^{N_{\rm m}}
    \omega^{k}(\theta_i) \,.
\end{equation}

\begin{figure}
    \centering
    \includegraphics[width=\columnwidth]{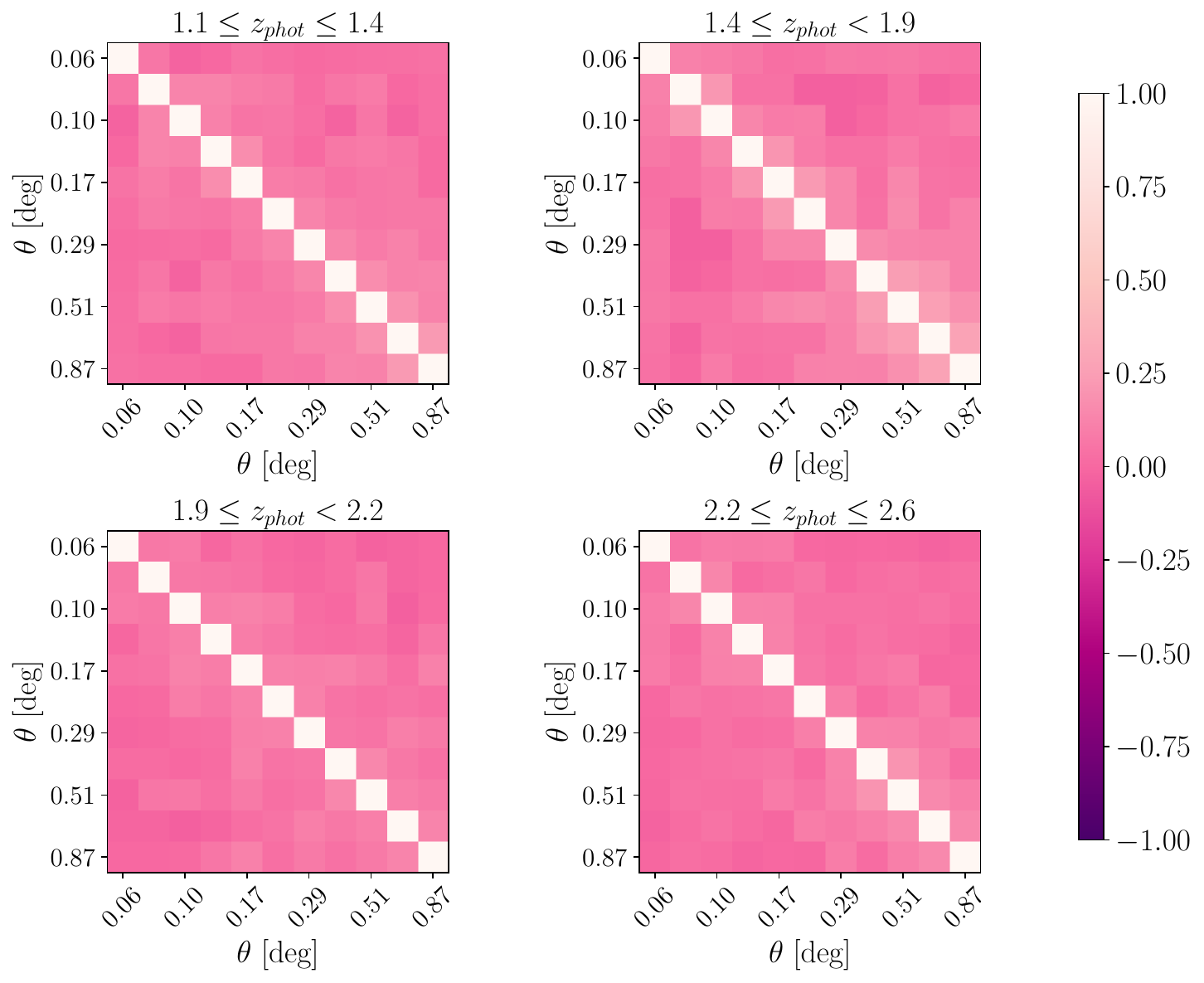}
    \caption{{Correlation matrices derived from the covariance matrices estimated from the $N_{\rm m} = 1,000$ mock catalogs for the four tomographic redshift bins. The color scale represents the correlation coefficient, defined as $R_{ij}=C_{ij}/\sqrt{C_{ii}C_{jj}}$.}}
    \label{fig:cov_matrix}
\end{figure}

The uncertainty of the 2PACF in each angular bin is then estimated from the square root of the diagonal elements of $\mathrm{C}_{ij}$. 
The corresponding correlation matrix is shown in Figure~\ref{fig:cov_matrix}. 
The covariance matrix of each interval is also used in the parameter estimation procedure through the $\chi^2$ minimization described in the following section.


\subsection{Theoretical Angular Correlation Function and Bias Estimation }\label{sec:ajuste}

Quasars are biased tracers of the large-scale matter distribution. In the linear regime,
their three-dimensional density contrast, $\delta_Q(\mathbf{x},z)$, is assumed to be
related to the underlying matter (dominated by dark matter) density contrast, $\delta_m(\mathbf{x},z)$, through a
linear bias parameter, $b(z)$, as commonly adopted in large-scale structure analyzes
\citep{Kaiser1984,Desjacques2018}. 
Under this assumption, the relation between the quasar and matter density contrast can be written as $\delta_Q(\mathbf{x},z)=b(z) \delta_m(\mathbf{x},z)$.

Assuming that the bias is scale-independent over the scales considered, the quasar and matter two-point correlation functions are related by $\xi_Q(r,z)=b^2(z) \xi_m(r,z)$ in three-dimensional space. 
Projecting this relation onto the celestial sphere, we can write
\begin{equation}
\omega_Q(\theta)=b^2(z)\omega_m(\theta) \,,
\label{eq:bias_scaling}
\end{equation}
where $\omega_Q(\theta)$ and $\omega_m(\theta)$ denote the angular two-point correlation functions of quasars and matter, respectively. Therefore, the quasar bias can be estimated by comparing the measured quasar 2PACF with the theoretical matter prediction.

The theoretical matter angular correlation function,
$\omega_m(\theta)$, is computed from the matter angular power spectrum,
$C_\ell^{mm}$, projected onto the celestial sphere.
The angular power spectrum is obtained from the projection of the
three-dimensional matter power spectrum, $P_m(k,z)$, along the line of sight.
Under the Limber approximation \citep{Limber1953},
which is accurate for sufficiently small angular scales, the projection becomes
\begin{equation}
C_{\ell}^{mm}
=
\int dz\,
\frac{H(z)}{c}
\frac{W^2(z)}{\chi^2(z)}
P_m\!\left(
k=\frac{\ell+1/2}{\chi(z)},z
\right) \,,
\label{eq:cell_limber}
\end{equation}
where $H(z)$ is the Hubble parameter,
$\chi(z)$ is the comoving radial distance, $c$ is the speed of light,
and $\ell$ is the angular multipole, approximately related to the angular scale by $\theta\simeq180^\circ/\ell$. 
The radial kernel is given by 
\begin{equation}
W(z) = \left(\frac{dN}{dz}\right)_{\rm norm} \,,
\label{eq:kernel}
\end{equation}
where the right hand side corresponds to the normalized redshift distribution of quasars in the respective tomographic bin. 
Ideally, this distribution should correspond to the true redshift distribution of the quasar population considered. However, since we do not have individual spectroscopic redshifts for the sample, we estimate this distribution from sampling the photometric redshift PDFs, as described in Subsection~\ref{sec:sampling}.

In the calculation of $C_\ell^{mm}$, we set the matter bias to $b_m=1$, such that the resulting angular power spectrum describes the projected clustering of the matter field itself, weighted only by the radial selection function. Finally, the theoretical angular correlation function is then obtained from the angular
power spectrum through the Legendre expansion
\begin{equation}
\omega_m(\theta)
=
\sum_{\ell}
\frac{2\ell+1}{4\pi}
C_\ell^{mm}
P_\ell(\cos\theta) \,,
\label{eq:wtheta_legendre}
\end{equation}
where $P_\ell(\cos\theta)$ are the Legendre polynomials.

Assuming that the linear bias is approximately constant within each tomographic bin,
Eq.~(\ref{eq:bias_scaling}) implies
\begin{equation}
\omega_{\rm obs}(\theta)
\simeq
b_{\rm eff}^{\,2}\,
\omega_m(\theta) \,,
\label{eq:beff_fit}
\end{equation}
so that the effective bias of each redshift bin is determined by fitting the
amplitude of the theoretical matter correlation function to the observed
angular correlation function. The bias is then estimated by minimizing the $\chi^2$ statistic, defined as
\begin{equation}\label{eq:chi2}
\chi^2 =
[\omega_{obs} -
\omega_{m}]^{\mathrm{T}}
\mathbf{C}^{-1}
[\omega_{obs} -
\omega_{m}] \,,
\end{equation}
where $\mathbf{C}$ is the covariance matrix of the measurements obtained in Section~\ref{sec:cov_matrix}.

The calculations assume a fiducial flat-$\Lambda$CDM cosmology, for which we adopt the cosmological parameters listed in Table~\ref{table:glass}, complemented by $\sigma_8=0.8120$ and $n_s=0.9649$, which are the parameters of the amplitude of
matter density fluctuations on a scale of $8h^{-1}$Mpc and spectral index, following the \textit{Planck} 2018 cosmological results \citep{planck2020}, and are performed using the open-source \texttt{Core Cosmology Library (CCL)}\footnote{\url{https://github.com/LSSTDESC/CCL}}\citep{Chisari2019}.

\section{Main Results} \label{sec:Results}

In this section, we present the main results of our tomographic clustering analysis. 
We first show the measured 2PACF for each tomographic redshift bin, then derive the corresponding effective bias values, and finally assess their consistency with theoretical predictions and previous studies.

\subsection{2PACF measurements in tomographic bins} \label{sec:results_2PACF}

The 2PACF was measured in four tomographic redshift intervals, defined by $1.1 \leq z_{\rm phot} < 1.4$, $1.4 \leq z_{\rm phot} < 1.9$, $1.9 \leq z_{\rm phot} < 2.2$, and $2.2 \leq z_{\rm phot} \leq 2.6$, using the sampling methodology, described in Section~\ref{sec:sampling}. In all bins, the LS estimator, Equation~(\ref{eq:2pacf}), was used, employing the random catalog constructed 
in Section~\ref{sec:random}, based on the same observational geometry as the selected sample.

\begin{figure*}
    \centering
    \includegraphics[width=\linewidth]{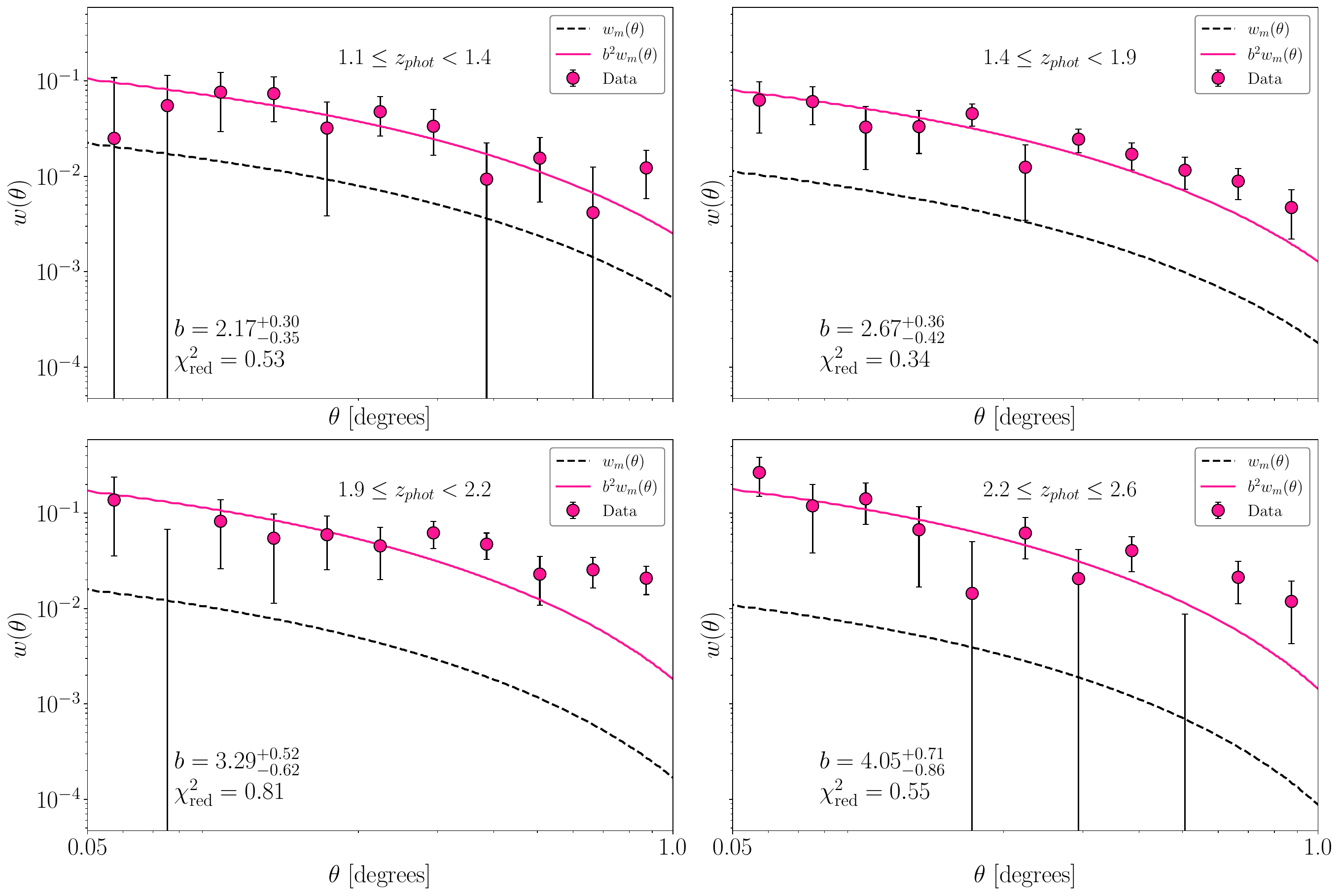}
    \caption{Measured angular two-point correlation functions, $\omega(\theta)$, of the quasar sample in the four tomographic photometric redshift bins. The deep pink data points with error bars correspond to the measured 2PACF. The black dashed curves show the theoretical matter two-point correlation functions $\omega_{\rm m}(\theta)$, with $b_m=1$, computed for the fiducial flat-$\Lambda$CDM cosmology using the sampled redshift distributions of each tomographic bin. The solid deep pink curves represent the best-fit models, $b^2\omega_{\rm m}(\theta)$, from which the effective linear bias was determined. The best-fit bias values and the reduced chi-square, $\chi_\nu^2$, are reported in each panel.}
    \label{fig:fits_omegas}
\end{figure*}

Figure~\ref{fig:fits_omegas} shows the 2PACF measurements for each tomographic bin, where the error bars associated with the measurements correspond to the square root of the diagonal elements of the covariance matrix (Subsection~\ref{sec:cov_matrix}). A positive signal is observed at the smallest angular separations, followed by a general gradual decrease in amplitude as $\theta$ increases, a behavior expected for the large-scale angular correlation function. 
Despite increased uncertainties in the high-redshift bins investigated, the clustering signal remains detectable across all intervals, allowing for the reliable determination of the effective bias in each tomographic bin.

\subsection{Effective Quasar Bias Estimation} \label{sec:effective_bias}

To estimate the effective linear bias, we compare the measured 2PACF from the data with the theoretical angular correlation function, derived in Section~\ref{sec:ajuste}, for matter (i.e., $b_m = 1$). 
Linear bias adjustment $b(z)$ was performed considering the angular clustering of quasars in the range $0.05^{\circ} \leq \theta \leq 1.0^{\circ}$, divided into 11 angular bins.

For each tomographic bin, the effective redshift, $z_{\rm eff}$ was calculated as the redshift weighted by the corresponding normalized redshift distribution, $dN/dz$, obtained from the Monte Carlo realizations described in Section~\ref{sec:sampling}. 
That is, the effective redshift is obtained with 
\begin{equation}
z_{\rm eff} =
\frac{
\displaystyle \int_{z_{\min}}^{z_{\max}}
z\,\frac{dN}{dz}\,{\rm d}z
}{
\displaystyle \int_{z_{\min}}^{z_{\max}}
\frac{dN}{dz}\,{\rm d}z
}.
\end{equation}

To estimate the bias parameter in each tomographic bin we adopted an approach based on minimizing the $\chi^2$ statistic. 
For each bin, the value of $b(z_{\rm eff})$ corresponding to the best-fit was determined by minimizing the statistic defined in Equation~(\ref{eq:chi2}), while its uncertainties were obtained from the profile of $\chi^2[b(z_{\rm eff})]$, considering the values that satisfy $\Delta\chi^2 = \chi^2[b(z_{\rm eff})]-\chi^2_{\rm min}=1$, corresponding to the $1\sigma$ confidence interval for a single free parameter.  We used the reduced chi-squared,  $\chi^2_{\rm red}$, defined as $\chi^2_{\rm red} \equiv \chi^2_{\rm min}/\nu$, where $\nu$ corresponds to the degrees of freedom, that is, the difference between the number of angular bins of $\omega_{\text{obs}}$ and the number of parameters freely fitted, to assess the quality of our fit. A summary of our results can be found in Table~\ref{table:summary_bias}. 

The 2PACF is measured directly from the quasar catalog, without assuming a cosmological model. However, the generation of simulated realizations requires fiducial cosmology, as does the comparison with theoretical predictions. Therefore, the values of $b_{Q}$ reported in this work should be understood as estimates relative to the prediction of the flat-$\Lambda$CDM model for the matter correlation function, where $b_m = 1$. 
A joint analysis of cosmological parameters and tracer bias could, in principle, explore alternative cosmological models, but would require additional observables to disentangle the effects of cosmology and tracer bias on the observed clustering amplitude \citep{Zheng2007}.

\begin{table}
\caption{Summary of bias fitting over $0.05 < \theta < 1.0$ deg$^{2}$ in the four redshift bins analyzed.}
\centering

\begin{tabular*}{\columnwidth}{@{\extracolsep{\fill}}c c c c}
\hline\hline
Interval & $z_{\text{eff}}$ & $b_\text{Q}$ & $\chi^{2}$(10 d.o.f.)\\
\hline
$1.1 \leq z < 1.4$ & $1.26$ & $2.17^{+0.30}_{-0.35}$ & $5.259$\\
$1.4 \leq z < 1.9$ & $1.63$ & $2.67^{+0.36}_{-0.42}$ & $3.361$\\
$1.9 \leq z < 2.2$ & $2.04$ & $3.29^{+0.52}_{-0.62}$ & $8.148$\\
$2.2 \leq z \leq 2.6$ & $2.38$ & $4.05^{+0.71}_{-0.86}$ & $5.519$\\
\hline
\end{tabular*}

\begin{minipage}{\columnwidth}
\vspace{0.2cm}
\justify
\textbf{Note.} Values of the effective redshift, $z_{\rm eff}$, the quasar bias parameter obtained from the best-fit, $b_{\rm Q}$, and the minimum value of $\chi^2$, with 10 degrees of freedom (d.o.f.), for each tomographic bin considered in the analysis.

\end{minipage}
\label{table:summary_bias}
\end{table}

\subsection{Discussion of the results}

In this section, we compare our results with measurements reported in the literature. 
However, as noted by~\citep{William2025}, comparisons between measurements should be made with caution, because different analyses adopt a distinct set of cosmological parameters for the fiducial cosmology, information that is then used to estimate uncertainties and to derive theoretical correlation functions.

\begin{figure}
\centering
\includegraphics[width=\columnwidth]{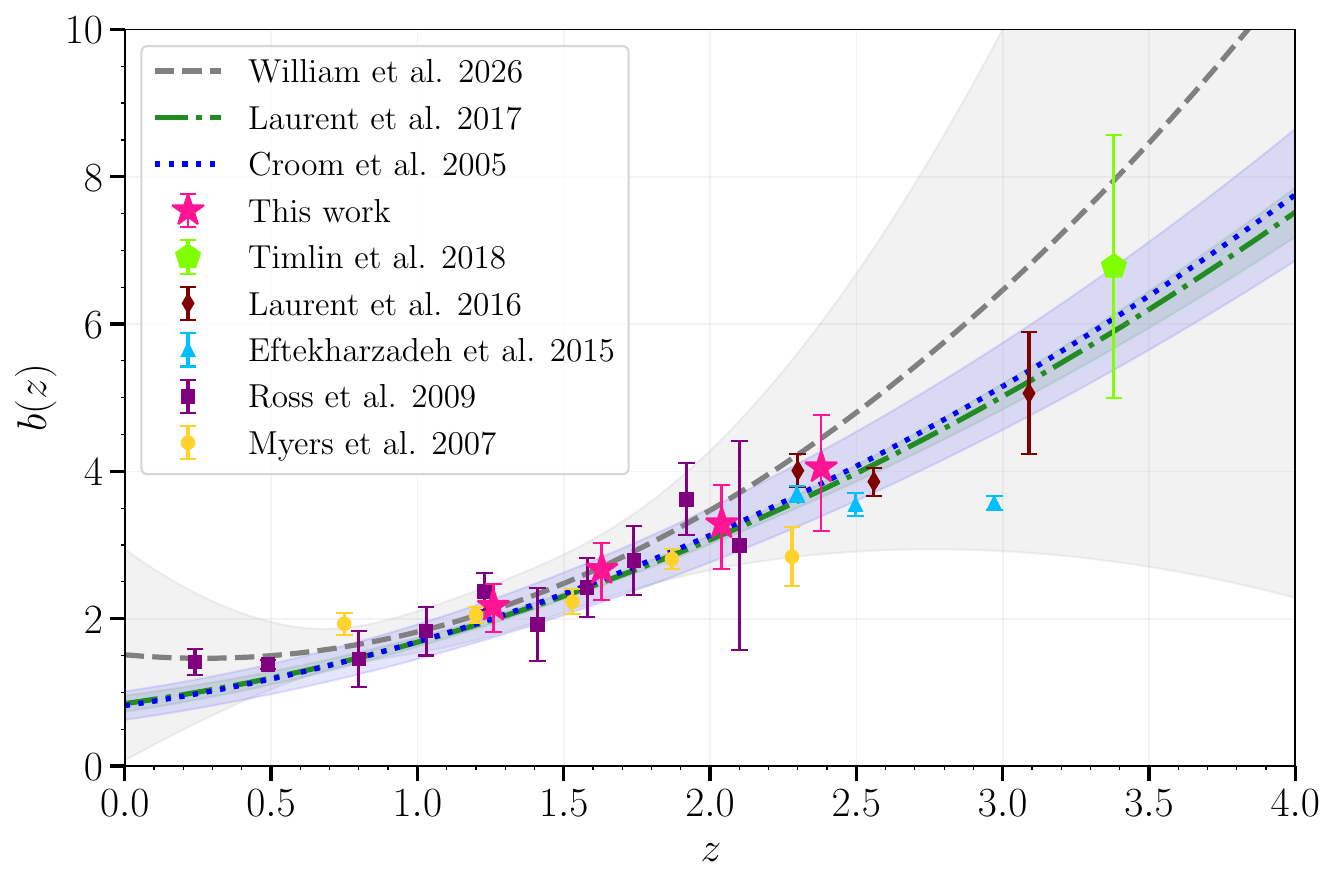}
\caption{The pink stars represent the measurements from this study for photometric quasars from the QuCatS catalog (S-PLUS DR4). The curves and markers correspond, respectively, to: the best-fit for quasar bias evolution from \cite{William2025} (gray dashed line), \cite{laurent2017} (dark green dashed-dotted line), \cite {Croom2005} (blue dotted line), \cite{Timlin2018} (light green diamond),  \cite{Laurent2016} (maroon diamond), \cite{Eftekharzadeh2015} (cyan triangles), \cite{Ross2009} (purple squares), and \cite{Myers2007} (gold circles).
}
\label{fig:literature}
\end{figure}

As displayed in Figure~\ref{fig:literature}, 
our four measurements agree within $1\sigma$ with the parameterizations in \cite{William2025, laurent2017, Croom2005}. 
It is worth noting that the different bias parameterizations exhibit quite distinct statistical behaviors. In \cite{William2025}, the polynomial parameterization $b(z)=b_0+b_1z+b_2z^2$ presents high uncertainties, especially at high redshifts, as well as strong correlations among the coefficients $(b_0, b_1, b_2)$, as evidenced by the reported covariance matrix, where the errors even exceed the fitted values themselves. This scenario likely stems from the lower statistical precision of the KiDS DR4 sample and the model's high sensitivity to the calibration of the redshift distribution employed. On the other hand, the parameterization proposed by \cite{laurent2017}, $b_Q(z)=\alpha\left[(1+z)^2-6.565\right]+\beta$, yields significantly smaller statistical errors and was constructed to ensure the absence of correlation between the parameters, i.e., $\rho_{\alpha,\beta}=0$. 
This feature provides the model with a more robust statistical constraint, allowing each parameter to be determined independently. In contrast, the model $b_Q(z)=b+a(1+z)^2$ from \cite{Croom2005}, while featuring parameters with relatively small individual uncertainties, suffers from strong anti-correlation between the coefficients. Thus, from a statistical standpoint, the \cite{laurent2017} parameterization offers the most stable and least degenerate description among the three approaches considered.

As a result of our analyses, we obtained four novel measurements of the function $b(z)$ consistent with the literature, which expand the available observational data. 
Although the sample on its own is insufficient to constrain parameterizations with 
two free parameters, it contributes to future joint analyzes with larger samples.

\section{Summary} \label{sec:summary}

In this work, we study a sample of quasars from the QuCatS value-added catalog, data obtained from the DR4 of the S-PLUS photometric narrow-band survey. 
We perform a tomographic analysis of quasar clustering using a sample of 23,402 cosmic objects, covering the range $1.1 \leq z_{\rm phot} \leq 2.6$. 
This sample was divided into four tomographic bins defined by $1.1 \leq z_{\rm phot} < 1.4$, $1.4 \leq z_{\rm phot} < 1.9$, $1.9 \leq z_{\rm phot} < 2.2$, and $2.2 \leq z_{\rm phot} \leq 2.6$, for which we calculated the angular two-point correlation function for each one. 
We produced $1,000$ lognormal mocks to estimate the covariance matrix used in the analysis. 
To propagate the uncertainties associated with photometric redshifts to our clustering measurements, we employ the inverse sampling technique \citep{ulisses2026}. 
Finally, we estimated the effective linear quasar bias for each tomographic bin by comparing the measured correlation functions with theoretical predictions for dark matter clustering assuming, as fiducial cosmology, the flat-$\Lambda$CDM cosmological model. 
We obtained $b_{Q}(z_{\rm eff}=1.26)=2.17^{+0.30}_{-0.35}$, $b_{Q}(z_{\rm eff}=1.63)=2.67^{+0.36}_{-0.42}$, $b_{Q}(z_{\rm eff}=2.04)=3.29^{+0.52}_{-0.62}$, and $b_{Q}(z_{\rm eff}=2.38)=4.05^{+0.71}_{-0.86}$.

Our results show an evolution of quasar clustering, specifically an increasing with redshift, consistent with the behavior 
observed considering other measurements reported in the literature, and the linear bias values obtained are consistent with previous measurements. Furthermore, the evolution of the bias is consistent with the $b(z)$ parameterization proposed by \citep{laurent2017}. 
Although uncertainties are higher in the high-redshift bins, these measurements remain valuable for characterizing quasar bias evolution and extending the observational mapping of large-scale structure at high redshifts. 
In particular, to the best of our knowledge, our results constitute the first cosmological study of the tomographic clustering of photometric quasars using S-PLUS data, contributing to the mapping of linear bias evolution across a redshift range where few measurements based on photometric samples currently exist. 
Our results also open up prospects for future analyses using new S-PLUS data releases which are expected to provide larger samples and improved calibrations of photometric redshift probability distributions, as well as for complementary studies using other wide field photometric surveys. 
\begin{acknowledgments}
  
ML thanks Coordenação de Aperfeiçoamento de Pessoal de Nível Superior (CAPES) for the financial support, and AB acknowledges Conselho Nacional de Desenvolvimento Científico e Tecnológico (CNPq) (Process No. 306340/2025-9) for the fellowship. FA thanks Fundação Carlos Chagas Filho de Amparo à Pesquisa do Estado do Rio de Janeiro (FAPERJ), Process No. SEI-260003/001221/2025, for the financial support. ML also thanks Lucas Knust for computational support and for providing the computational resources used in this work.

\end{acknowledgments}


\software{
\texttt{NumPy}~\citep{harris2020},
\texttt{SciPy}~\citep{Virtanen2020},
\texttt{Astropy}~\citep{astropy:2013, astropy:2018, astropy:2022},
\texttt{Matplotlib}~\citep{Hunter:2007},
\texttt{healpy}~\citep{Zonca2019, gorski2005},
\texttt{PyCCL}~\citep{Chisari2019},
\texttt{CAMB}~\citep{camb2000, Lewis:2026mif},
\texttt{GLASS}~\citep{tessore2023},
\texttt{TreeCorr}~\citep{jarvis2004},
\texttt{TOPCAT}~\citep{taylor2005},
\texttt{ChatGPT}~\citep{OpenAI2026},
\texttt{Claude}~\citep{Anthropic2026}.
}

\bibliographystyle{apsrev4-1}
\bibliography{main}

\end{document}